\documentclass[a4paper,aps,pra,
preprintnumbers]{revtex4}

\usepackage{graphics,graphicx,epsfig}

\usepackage[centertags]{amsmath}
\usepackage{amsfonts}
\usepackage{amssymb}

\usepackage[centertags]{amsmath}
\usepackage{amsfonts}
\usepackage{amssymb}
\usepackage{amsthm}
\newcommand{\beq}{\begin{equation}}
\newcommand{\eeq}{\end{equation}}
\newcommand{\beqa}{\begin{eqnarray}}
\newcommand{\eeqa}{\end{eqnarray}}

\newcommand{\ket} [1] {\vert #1 \rangle}
\newcommand{\bra} [1] {\langle #1 \vert}

\newcommand{\proj}[1]{\ket{#1}\bra{#1}}
\newcommand{\mean}[1]{\langle #1 \rangle}

\begin{document}

\sloppy
\begin{abstract}
Large number of multimode entangled states of light generated in
down conversion processes belongs to a collection which is natural
generalization of the $W$ class. A brief overview of these states,
schemes for their preparation, experimental implementations and
possible applications are presented.
\end{abstract}
\title{On multiparticle $W$ states, their implementations and
applications in quantum information problems}
\author{V.N. Gorbachev, A.I. Trubilko}.
\affiliation {Laboratory for quantum information  $\&$
computation,
University of AeroSpase Instrumentation,\\
St-Petersburg, 190000, Bolshaia Morskaia 67, Russia.}

\maketitle
\section{Introduction}

Entangled states have been generated in many experiments  and
their applications in quantum communications have been
demonstrated (see for example \cite{EntExp}). Now great efforts
are concentrated on investigation of multiparticle entanglement
with its promising features which are interesting in
decoherence-free quantum informational processing, advanced
multiparty quantum communications and others.

 When considering implementation of the multiparticle
entangled states (MES), one finds that most of them have been
generated in optics experiments. In a  typical  optical scheme
photons of the source are distributed to output modes by linear
lossless elements. If the configuration is symmetric and a photon
enters input, then it can be found on any of the outputs. By this
way the multimode light is achieved, its  state belongs to the
$W$-class introduced by Cirac et al \cite{Cirac W}. Varying
 configuration of the scheme as well as using the  source of  polarized
photons a more extensive collection of MES than $W$-class arises.
Particularly  it includes the Dicke states and can be studied from
 point of view of quantum information theory (QIT)
without referring to any physical system. Analysis of common
features gives us an  answer  how to manipulate MES in optimal way
and which of informational tasks can be done using a given
entangled state. In respect to physics MES are natural states of
the multiparticle systems. So if any $m$ particles of a large
ensemble can be excited then all possibilities result in the
considered MES. Indeed entangled states of two macroscopic atomic
ensembles have been demonstrated experimentally by Polzik et al
\cite{Polzik Nature}.

It is important to know the common features of MES because  it
determines how to detect them in experiment. In QIT there is a set
of criteria of entanglement which require the knowledge of the
state or its entropy also Bell inequalities are often discussed.
But exploiting such criteria in experiment is a hard problem.
 However there are specific witness observables which
expectation values indicate entanglement. Experimental
implementation of the witness observable for polarized light have
been demonstrated by Weinfurter et all \cite{Weinfurter Witness}.
The estimation of an unknown state can be made by quantum
tomography. Indeed this method has been used by Roos et all
\cite{IonExpW} in experiment with trapped ions. Another way is
measuring  a such operator which eigenvector is the desired
entangled state. Then its eigenvalue and variance indicate
entanglement. In fact some members of  the $W$-class are reduced
to the Dicke states which are eigenvectors of two collective
operators $J^{2}$ and $J_{3}$ \cite{Dicke}. In optics
implementation the spin variables can be associated with
polarization of light and one can measure, for example, variances
of these operators which describe noise of light. For eigenvectors
these variances are equal to zero and it means that there is no
noise. More precisely in such measurement the shot noise of light
is suppressed bellow standard quantum limit given by the coherent
state.

 Quantum correlations
are fragile and easily  destroyed with environment nevertheless
several MES have immunity to decoherence and they are robust to
loss of particles. These features are attractive for quantum
communications, but their exploiting is a hard problem. For
example, many attempts have been made to introduce $W$ state
instead of the EPR pair in the standard teleportation protocol
proposed by Bennett et al \cite{Bennett tel}. But most of these
proposals results in conditional teleportation, when the task is
accomplished with a probability.
However several of the $W$-states can be suitable as a quantum
channel for dense coding and the unconditional teleportation of
entangled states
also for the problem of secrete sharing
and other tasks.

The main aim of this work is to consider properties,
implementations and applications of the collection of MES, which
are  simple generalization of the $W$ class. For a particular case
of three-particle $GHZ$ and $W$ states transformations between
them, three-party quantum communications protocols for secrete
sharing and splitting of quantum information with $GHZ$ have been
discussed by Karlsson et al \cite{Karlsson Rew 3W 3GHZ}.
In our work using the standard approaches of QIT we pay attention
to physical features of MES.

This  paper is organized as follows. First we describe
tree-particle $W$ state, which is non-equivalent to $GHZ$ and
robust to loss  of particle. Then more general states are
introduced and their connection with the Dicke states together
with their properties, measures of entanglement and witness
observables are considered. Next we examine several proposals for
generating MES of atoms and light and overview experimental
implementations of the three and four photon $W$ states. Finally
several protocols are discussed.

\section{Properties}
\subsection{Three-qubit GHZ and W states}

\subsubsection{Classification of states in LOCC.}
There are strong definitions for the three-particle $W$ states
introduced by Cirac et al \cite{Cirac W}. These definitions are
based on a set of transformations known as LOCC (Local Operations
and Quantum Communication). We will use a simple notation, that
operators are local if $U\ket{\varphi}_{AB}=A\otimes
B\ket{\varphi}_{AB}$, where operator $A$ acts on the particle $A$
and don't affect to $B$ and so on. Following this point one finds,
that the two-mode hamiltonian described a parametric down
conversion source (PDC) $H=ik\hbar(a^{\dagger}b^{\dagger}-ab)$ is
an example of non-local operator. PDC generates entangled states
but they can't be created by LOCC. This is a general property of
entanglement.
\\
How to compare one state with another? From  physical reasons the
answer is clear. If a physical system is prepared in different
states, then by measuring of observables, one can distinct them in
principle. In QIT, which operates the logical states without
referring to any particular physical system, two states are
identical, if they can be obtained from each other with certainty
by LOCC.
Particularly, it means, that parties can use these two states for
the same task \cite{C.H. Bennett9511030}.
\\
In this approach there are two classes of irreducible tripartite
entanglement, namely either GHZ states \cite{Greenberger Bell's
theorem}
\begin{eqnarray}
  \ket{GHZ}=1/\sqrt{2}(\ket{000}+\ket{111}),
\end{eqnarray}
and
\begin{eqnarray}
\label{W}
  \ket{W}=1/\sqrt{3}(\ket{100}+\ket{010}+\ket{001}),
\end{eqnarray}
known as W-state in QIT. Here $\ket{0},\ket{1}$ are states of a
two-level system, or qubit (quantum bit). One finds, that
$\ket{GHZ}$ can't be converted into $\ket{W}$ by LOCC, but they
are entangled because of their wave functions are not factorized
into product of three particles. For general case the $W$-class
introduced by Cirac has the form
$\varphi_{W}=a\ket{000}+b\ket{100}+c\ket{010}+d\ket{001}$, where
$a^{2}+b^{2}+c^{2}+d^{2}=1$.

\subsubsection{Robustness}

Some differences between $\ket{GHZ}$ and $\ket{W}$ are clear
without LOCC.
\\
Consider any two of three particles, say 1 and 2, which density
matrix is $\rho(12)=Tr_{3}\rho(123)$. If one of the particles is
traced out, in QIT  it means a loss of particle. In fact three
parties share particles 1,2 and 3 in entangled state and one of
them decides not to cooperate with other two. Can the remainder
two parties accomplish the task? Answer depends on the robustness
of the state. For $GHZ$
\begin{eqnarray}
  \rho_{GHZ}(12)=1/2(\proj{00}+\proj{11}),
\end{eqnarray}
for $W$
\begin{eqnarray}
\label{RRR}
  \rho_{W}(12)=1/3\proj{00}+2/3\proj{\Psi^{\dagger}},
\end{eqnarray}
where $\Psi^{\dagger}=1/\sqrt{2}(\ket{01}+\ket{10})$. The main
difference is that the density matrix $\rho_{W}(12)$ has
non-diagonal elements or coherence $\ket{01}\bra{10}$. This is a
reason of entanglement of $\rho_{W}(12)$ in contrast to
$\rho_{GHZ}(12)$.
\\
In more detail. To analyze the mixed states it needs to introduce
 criteria of inseparability, that are generalizations of the
 non-factorizability of the wave function. A criterion of Werner \cite{Werner} tells, that
 state is separable or classically correlated or non-entangled
if its density matrix has the form
\begin{eqnarray}
  \rho(12)=\sum_{k}\lambda_{k}\rho(1)_{k}\otimes\rho_{k}(2),
\end{eqnarray}
where $\sum_{k}\lambda_{k}=1$ and all $\lambda_{k}\in [0,1]$. Then
one finds that $\rho_{GHZ}(12)$ is classically correlated. Next
criterion is necessary and sufficient to establish inseparability
of $\rho_{W}(12)$. It tells, that the state which dimension of
Hilbert space is $2\times 2$ or $2\times 3$ is inseparable, if any
of eigenvalues of the partially transposed density matrix is
negative \cite{Peres}, \cite{Horodecki Criterium}. The partially
transposed density matrix, say over particle 1, reads
$\rho^{T_{1}}(12)_{W}=1/3\proj{00}+2/3(\proj{01}+\proj{10}+
\ket{11}\bra{00}+\ket{00}\bra{11})$, it has a negative eigenvalue,
then $\rho_{W}(12)$ is inseparable or entangled. In this case
quantum correlations between the remainder two particles survive
after tracing and this is  robustness to particle loss.
\\
\subsection{Several generalizations}

\subsubsection{Multiparticle $W$, $ZSA$ and other states}

 A simple generalizations is possible by introducing the
n particle states of the form
\begin{eqnarray}
\label{eta}
  \eta_{n}(1)=q_{1}\ket{10,\dots,0}+q_{2}\ket{01,\dots,0}+\dots
  +q_{n}\ket{00,\dots,1}),
\end{eqnarray}
where $\sum_{k}|q_{k}|^{2}=1$. In particular case
\begin{eqnarray}
  q_{1}+\dots q_{n}=0,
\end{eqnarray}
one finds the zero sum amplitude (ZSA) states introduced by Pati
\cite{Pati}. They are not equivalent to $GHZ$ under LOCC. If all
coefficients in $\eta_{n}(1)$ are equal there is a totally
symmetric wave function
\begin{eqnarray}
  W_{n}=1/\sqrt{n}(\ket{10,\dots,0}+\ket{01,\dots,0}+\dots
  +\ket{00,\dots,1}),
\end{eqnarray}
that is known in QIT as multiparticle W-states \cite{Cirac W}.
Such symmetric vector describes an ensemble of two-level physical
systems, qubits, where only $m=1$ particle from $n$ is excited.
When $m\geq 1$ the symmetrized states has the form
\begin{eqnarray}
\label{Sym}
  \ket{m;n}=1/Q\sum_{z}P_{z}\ket{\underbrace{1,\dots
  1}_{m},\underbrace{0,\dots,0}_{n-m}},
\end{eqnarray}
where $P_{z}$ is one from $C^{n}_{m}=n!/(m!(n-m)!)$
distinguishable permutations of particles, $Q=\sqrt{C^{n}_{m}}$.
All states in the superposition (\ref{Sym}) have  equal weight
$Q$, when their weights are different  the more general
$\eta_{mn}$ states may be found, which are a natural
generalization of the $W$-class. The introduced collection of
$\eta_{mn}$ contains $W$, $ZSA$ and symmetric states $\ket{m;n}$.
\\
These states seem to be natural for multiparticle systems and can
be generated in many physical processes. When considering problem
of interaction between atoms and light, for example, the
collective atomic operators are introduced. For two-level
identical atoms they read
\begin{eqnarray}
 \label{atS}
 S_{xy}=\sum_{a}\ket{x}_{a}\bra{y},
\end{eqnarray}
where $\ket{x}_{a}\bra{y}, x,y=0,1$ is operator of a single atom
$a$ and 0,1 label lower and upper level. There is a representation
for $\ket{m;n}$
\begin{eqnarray}
  S_{10}^{m}\ket{0,\dots,0}=m!Q\ket{m;n}.
\end{eqnarray}
It tells, that any symmetric states $\ket{n;m}$, particulary $W$
ones, can be produced in any process of collective interaction
between atoms and light or other system. There is a simple
physical reason for it. When each of $m\leq n$ identical atoms
absorb a photon, then all possibilities results in superposition
$\ket{m;n}$, when atoms are distinguished more general state
$\eta_{nm}$ arises.

\subsubsection{Connection with Dicke states}
Some members of the $W$ class can be reduced to the Dicke states.
Let introduce collective operators $J_{k}, k=1,2,3$ and
$J^{2}=J_{1}^{2}+J_{2}^{2}+J_{3}^{2}$, that obey the commutation
relations of the momentum operators
\begin{eqnarray}
[J_{j};J_{k}]=i\epsilon_{jkl}J_{l}.
\end{eqnarray}
The Dicke states are defined as eigenvectors of two operators
$J^{2}$ and $J_{3}$ \cite{Dicke}
\begin{eqnarray}
  J^{2}\ket{jl}=j(j+1)\ket{jl},~~~J_{3}\ket{jl}=l\ket{jl},
\end{eqnarray}
where $|l|\leq j$, $\max j=n/2$, $n$ is number of the particles
and $l=[(n-m)-m)]/2$ is difference between the numbers of
non-excited and excited particles. Using (\ref{atS}), we have a
representation
\begin{eqnarray}
\label{rrr}
J_{1}=(S_{10}+S_{01})/2,~~J_{2}=i(S_{10}-S_{01})/2,~~J_{3}=(S_{00}-S_{11})/2.
\end{eqnarray}
Note, that in (\ref{rrr}) the vectors $\ket{0},\ket{1}$ however
can be considered as the Fock state of light.
\\
All symmetric states have the form $\ket{j=n/2,l}$ and
\begin{eqnarray}
  W_{n}=\ket{j=n/2,~l=n/2-1}.
\end{eqnarray}
For ZSA states
\begin{eqnarray}
  \eta_{n}=\ket{j=n/2-1,~ l=n/2-1}.
\end{eqnarray}
As result $W$ and $ZSA$ are eigenvectors of $J^{2}$ and $J_{3}$
and belong to family of the Dicke states.

\subsubsection{Symmetry and Decoherence-Free states}

When the symmetry of state is conserved in a  physical processes
one finds, for example, that the antisymmetric two-particle wave
function $\Psi^{-}=(1/\sqrt{2})(\ket{01}-\ket{10})$ can't be
transformed into product $\ket{00}$.
 Such entanglement is therefore
decoherence-free (DF). A state is DF if it is invariant under some
unitary transformation, described an collective interaction with
noisy environment. It is interesting for protection of information
by the noiseless quantum code, that has been considered by Zanardi
et al \cite{Zanardi Noiseless}.
\\
Let $\Psi^{-}$ be the state of two atoms, which interact with its
thermostat, then
\begin{eqnarray}
\label{dec}
 S_{01}\Psi^{-}=0.
\end{eqnarray}
Formally $\Psi^{-}$ looks as a vacuum and has immunity to the
spontaneous decay. This state can be storage in a collective
thermostat which has been considered by Basharov \cite{Askhat}. It
is described by equation of the Lindblad form\begin{eqnarray}
\dot{\rho}=-\gamma[R^{\dagger}R\rho-R\rho R^{\dagger}+h.c.],
\end{eqnarray}
where $\gamma$ is a decay rate and $R=S_{01}$ is a collective
operator.
\\
To preserve entanglement, in QIT a class of DF states has been
introduced \cite{Palma}, \cite{Zanardi Noiseless}, \cite{Lidar},
\cite{Kempe}. For two particles there is only one DF state
$\Psi^{-}$. In the case of four particles there are two DF states,
which are interesting for applications. They are a product of
singlets $\Psi^{-}$
\begin{eqnarray}
\label{DF1}
 \Phi_{0}=\ket{\Psi^{-}}\ket{\Psi^{-}}&&
\end{eqnarray}
and an  orthogonal to $\Phi_{0}$ vector, introduced by Kempe
\cite{Kempe}
\begin{eqnarray}
\label{DF2}
 \Psi_{1}=(1/\sqrt{3})(\ket{0011}+\ket{1100}-
 \ket{\Psi^{+}}\ket{\Psi^{+}}).
\end{eqnarray}
To protect quantum information, the logical qubit
$\alpha\ket{0}+\beta\ket{1}$ can be encoded in to superposition
$\alpha\Psi_{0}+\beta\Psi_{1}$, which is DF state and immune
against noise. These  two DF states have been generated
experimentally by Weifurter et al \cite{Weifurter DF} to
demonstrate DF quantum information processing.
\\
Several examples of DF states of physical systems can be
introduced by a simple generalization of (\ref{dec}). Consider
$\eta_{n}(1)$, where  logical qubits are implemented by two-level
atoms or by  modes of light in the Fock state with $0$ and $1$
photon. Next observation is true. There is a collective operator
$R$, for which
\begin{eqnarray}
\label{decN} R\eta_{n}=\sum_{k}q_{k}\ket{0}.
\end{eqnarray}
It can be chosen in the form $R=S_{10}$ or $R=\sum_{k}a_{k}$,
where $a_{k}$ is annihilation operator of the light mode $k$. If
$\sum_{k}q_{k}=0$, one finds ZSA states, which are robust to
Decoherence like  $\Psi^{-}$. So that either atomic ensemble or
light is prepared in ZSA state it may conserve its quantum
correlations under a collective noisy environment.

\subsection{Entanglement of multiparticle $W$ states}

\subsubsection{Entanglement and its measures}

Entanglement of multiparticle system can depend from the number of
particles. The reason is that if single excitation is distributed
into a large number of particles then total state is closer to
unexcited or vacuum state.
\\
In more detail. Introducing a  density matrix of $n$ particles
$\rho(n)= \proj{W_{n}}$ and considering a  state of any $s\leq n$
particles we have
\begin{eqnarray}
\label{rho s}
  \rho(s\leq n)=(s/n)\proj{W_{s}}+(1-s/n)\proj{0}.
\end{eqnarray}
For $s=2$ the Peres-Horodecki inseparability criterion  can be
applied. The  eigenvalues of the partially transposed density
matrix are $\lambda=\{1/n;1/n;(n-2)[1\pm\sqrt{1+4/(n-2)^{2}}]/2n$,
where one of them is negative. Thus the state is inseparable or
entangled. However in the limit of $n\to\infty$ entanglement
vanishes: $\lambda=\{0;0;0;1\}$, that is in agreement with
(\ref{rho s}), from which it follows, that $\rho(s\leq
n)\approx\proj{0}$.
\\
The problem whether a given  multiparticle state is entangled is
hard because of the calculation difficulty increases exponentially
with number of particles.
 There is
no necessary and sufficient operational criterion and various
measures of entanglement are used. Several common measures are
entanglement entropy, entanglement formation and negativity.
\\
Entanglement entropy $E(\ket{\Psi}_{AB})$ of a pure state and a
partition for the system $A,B$ is defined as
$E(\ket{\Psi}_{AB})=S(\rho_{A})=S(\rho_{B})$, where
$S(\rho)=-Tr(\rho\log \rho)$ is von Neuman entropy and
$\rho_{A}=Tr_{B}\proj{\Psi_{AB}}$ is the reduced density matrix.
For product states entanglement entropy is zero. It has its
maximum $\log dim(A)$, given for a partition with dimension
$dim(A)=d_{A},dim(B)=d_{B}$ and $d_{A}<d_{B}$. A state that
achieves this maximum is maximally entangled:
$\Psi_{AB}=1/\sqrt{d_{A}}(\ket{0,0}+\ket{1,1}+
\dots+\ket{d_{A}-1,d_{A}-1})_{AB}$. For EPR pair of the form
$\ket{\varphi}=\alpha\ket{00}+\beta\ket{11}$ we have
$E(\ket{\varphi})=H(p)$, where  $H(p)=-p\log p-(1-p)\log (1-p)$ is
a function of entropy, well known in the theory of information,
$p=|\alpha|^{2}=1-|\beta|^{2}$. If $p=1/2$ the function $H(p)$ has
its maximum corresponding to maximal entanglement. Using the
presented definition the entanglement entropy of the $W$-state
obtained from (\ref{rho s}) has the form of $H(p)$:
\begin{eqnarray}
 E(\ket{W_{n}})=-(s/n)\log (s/n)-(1-s/n)\log (1-s/n),
\end{eqnarray}
where $A$ and $B$ are subsystems of $s$ and $n-s$ particles. If
$s=n/2$, then $A$ and $B$ have the same number of particles and
their entanglement achieves maximum.
\\
Entanglement of formation is reduced to entanglement entropy for
pure states and defined as
$E_{F}(\rho_{AB})=\min_{\{p_{i}\ket{\psi_{i}}_{AB}\}}
\sum_{i}p_{i}E(\ket{\psi_{i}}_{AB})$,  where
$\rho_{AB}=\sum_{i}p_{i}\proj{\psi_{i}}$.
\\
The logarithmic negativity is defined as the absolute sum of the
negative eigenvalues of the partial transpose with respect to $A$
of density matrix $\rho_{AB}$. So
$N(\rho_{AB})=\sum_{i}(|\lambda_{i}|-\lambda_{i})/2$. Negativity
may also disagree with other measures for the so-called positive
partial transposed entangled states, which negativity is zero. It
demonstrates, that to analyze multiparticle entanglement it needs
various measures and criteria.
\\
There is a criterion, that can be interesting from the
experimental point of view. It is based on single-particle
measurement. The persistency of entanglement is defined as the
minimum number of single-particle measurements $M$ such that, for
all measurement outcomes, the state is completely disentangled
(separable) \cite{BriegelRaussendorf}. Let $\ket{0},\ket{1}$ be a
basis of the measurement.  Then for GHZ $M=1$, but for $W_{n}$ it
needs $M=n-1$ measurements to obtain a separable state.

\subsubsection{Witness}

In practice exploiting of the above measures is a hard problem and
any recipes adjusted to observables seem to be more appropriate.
\\
To detect various forms of multipatite correlations witness have
been introduced \cite{Horodecki Criterium}, \cite{Terhal}. A
witness of $n$- particle entanglement is an observable, which
value on state with $n-1$ partite entanglement is positive and
negative on some $n$-partite entangled state.
\\
A witness operator for the three-particle $W$ state reads
\begin{eqnarray}
 W^{(1)}_{W}=2/3-\proj{W}.&&
\end{eqnarray}
In accordance with definition $Tr[W^{(1)}_{W}\proj{W}]=-1/3$ and
$Tr[W^{(1)}_{W}\rho_{W}(12)]=1/9$, where two particle density
matrix $\rho_{W}(12)$ is given by (\ref{RRR}). This witness has
positive expectation value on biseparable and fully separable
states. Then it detects all tripartite entangled states of $W$ and
$GHZ$ classes, which can be distinguished by the second witness
operator
\begin{eqnarray}
 W^{(2)}_{W}=1/2-\proj{GHZ'},
\end{eqnarray}
where $\ket{GHZ'}$ is obtained from $\ket{GHZ}$ by replacing
$\ket{x}\to ((-1)^{x}\ket{0}+i\ket{1})/\sqrt{2}$, $x=0,1$.
\\
These witnesses can be observed in experiment. Using the Pauli
matrixes one finds
\begin{eqnarray}
\label{Ww1}
 \nonumber
W^{(1)}_{W}= 1/24[17+7\sigma_{z}^{\otimes
3}+3(\sigma_{z}\otimes1\otimes 1 +1\otimes \sigma_{z}\otimes
1+1\otimes1\otimes\sigma_{z})+5 (\sigma_{z}^{\otimes 2}\otimes
1 +\sigma_{z}\otimes 1\otimes\sigma_{z}+1\otimes\sigma_{z}^{\otimes 2})&&\\
\nonumber -(1+\sigma_{z}+\sigma_{x})^{\otimes 3}-
(1+\sigma_{z}-\sigma_{x})^{\otimes 3}-
(1+\sigma_{z}+\sigma_{y})^{\otimes 3}
-(1+\sigma_{z}-\sigma_{y})^{\otimes 3}].&&
\end{eqnarray}
In optics implementations two qubit states can be presented by the
polarization of the photons with horizontal $H$ and vertical $V$
linear polarization. Then it needs a set of polarization analyzers
to measure the witness operators \cite{Weifurter Witness}. Being
non-universal measure, witnesses provide the sufficient criteria.
\\
Another way of testing the $W$ entanglement is to measure the
operators $J^{2}$ and $J_{3}$ whose eigenvectors are $W$ and $ZSA$
states. In the representation given by (\ref{rrr}) the spin
variables can be associated with the $H$ and $V$ polarized
photons. Then by measuring the polarization of photons one can get
the expectation values of $J^{2}$ and $J_{3}$, whose variances are
zero for $W$ and $ZSA$ states. These expectation values and its
variances indicate the entanglement.
\section{Schemes for generation.}

\subsection{Atomic systems}
\subsubsection{Schemes}
There are several proposals on generating of $W$ states in Cavity
Quantum Electrodynamics (QED) and Raman interaction between
three-level atoms and the high-Q cavity modes. These models seem
to be attractive but they often neglect all relaxations processes
that is a hard problem in its implementation. Some principal
features can be demonstrated by considering a more simple model of
two-level atoms in free space.
\\
Interaction between an ensemble of two-level atoms and light can
be described by the usual Hamiltonian
\begin{eqnarray}
\label{hHH} H=i\hbar (S_{10}B-S_{01}B^{\dagger}),
\end{eqnarray}
where $B$ is a field operator. This Hamiltonian allows to examine
various processes of the one-photon interaction, for which $B=ga$,
Raman type scattering of two modes $a$ and $b$, when
$B=fa^{\dagger}b$, where $g,f$ are coupling constants. In the
model given by (\ref{hHH}) there are some physical reasons, that
result in $W$ states. Without relaxation one finds integral of
motion conserving the total number of excitations $m$. For
example, if all atoms are in their ground states and light in the
Fock state with one photon only, then $m=1$. In the case of
one-photon interaction the integral has the form
 $I=a^{\dagger}a-(1/2)[S_{00}-S_{11}].$
Then during evolution atoms and light exchange the excitation. As
result  $m$ is distributed into atomic ensemble  or into the light
modes and  the symmetric Dicke states $\ket{m,n}$ particularly
$W_{n}$ states of either atoms or modes are generated.
\\
Generally the multiparticle model given by (\ref{hHH}) is not
integrable. But for particular cases simple exact solutions can be
found. The reason is that in the symmetry-preserving interaction
only a part of states from the total Hilbert space is involved in
evolution. Then one can get simple analytic solutions, if the
problem includes a small number of $m$. In the case of Raman type
interaction we have \cite{Gorbachev parametric type interaction}
\begin{eqnarray}
\label{RI} \nonumber
 (\alpha\ket{01}+\beta\ket{10})_{ab}\otimes\ket{m;n}\to &&\\
\nonumber
 \alpha\{\cos\theta_{m}\ket{01}\otimes\ket{m;n}+
 \sin\theta_{m}
\ket{10}\otimes\ket{m+1;n}
 \}&&\\
 +\beta\{-\sin\theta'_{m}\ket{01}\otimes\ket{m-1;n}+\cos
 \theta'_{m}\ket{10}\otimes\ket{m;n}\},
\end{eqnarray}
where $|\alpha|^{2}+|\beta|^{2}=1$,
$\theta_{m}=tf\sqrt{(m+1)(n-m)}$, $\theta'_{m}=\theta_{m-1}$. This
example shows evolution of the totally symmetric initial state of
atoms $\ket{m;n}$ and entangled state
$\alpha\ket{01}+\beta\ket{10}$ of two modes. Equation (\ref{RI})
describes the next processes: 1/ generation of atomic $W$
entanglement $\ket{0;n}\to\ket{1;n}=W_{n}$, 2/ transformation of
the symmetric states $\ket{m;n}\to\ket{m\pm 1;n}$, 3/ entanglement
swapping, when the light state is transformed into atoms and back.
\\
Preparation of the $W$ and $GHZ$ atomic states in Raman type
interaction has been considered by Agarwal et al \cite{Agarwal}.
In this work the numerical analyze of the analytic solutions has
been presented.
\\
If an excited atom interacts with three or more cavity modes, it
can emits a photon into one of them, then the $W$ state of light
may be achieved \cite{Zhuo-Liang Cao W state via cavity QED}.
\\
Measurement is another way for  preparation of a physical system
in a given state, but it can be done with some probability. If
atoms interact with cavity modes, then by detecting an  output
photon the atomic $W$ and Dicke states may be achieved
\cite{Yun-Feng Xiao}.
\\
Interaction between atoms and light can produce  entanglement
between them. Such systems are useful for preparing  entangled
state of  atomic ensembles, when a projective measurement on
photons is performed.
 By this way $W$ states of atomic ensembles can be achieved
\cite{Gorbachev entangled W-states atom},  they have a hierarchic
organization being consisting of ensembles each of which is in the
$W$ state.
\\
Indeed the Heisenberg model was used to produce three-atom or
four-atom $W$ state in Ref. \cite{X. Wang}.
\subsubsection{Experiment with trapped ions}
Three qubit W and GHZ states of trapped ions have been generated
experimentally by Roos et all \cite{IonExpW} and conditional
operations for readout of an individual qubit have been
implemented.\\
In this experiment qubits are encoded in the ground and metastable
states $D$ and $S$ of the $^{40}Ca^{+}$ ion. A laser pulse can
rotate each ion
\begin{eqnarray}
R(\theta,\phi)=\exp[i\theta/2(e^{i\phi}\sigma^{+}+e^{-i\phi}\sigma^{-})],
\end{eqnarray}
that results in transitions between levels $D$ and $S$, where
$\sigma^{+}=\ket{S}\bra{D}$. By this way the $\pi/2$ pulse, for
which $R(\pi/2,0)$, creates a superposition
$1/\sqrt{2}(\ket{S}+i\ket{D})$, if initially ion is in $S$ state.
When ions are trapped in a linear Pauli trap each of them can
interact with vibrational mode due from motion
\begin{eqnarray}
R^{+}(\theta,\phi)=\exp[i\theta/2(e^{i\phi}\sigma^{+}b^{\dagger}+e^{-i\phi}\sigma^{-}b)],
\end{eqnarray}
where $b,b^{\dagger}$ are photon operators of the mode. Both
operations $R$ and $R^{+}$ were implemented experimentally. Using
them one can entangle ions with vibrational mode, rotate each ion,
map the state of modes into ions and other. As result
 from the initial state of the trapped
ions $\ket{SSS}$ the desired entanglement can be prepared.
\\
Indeed evolution given by $R$ and $R^{\dagger}$ has the same form,
when only one photon of vibrational mode is involved. Let
$\ket{0}=\ket{S}$ or $\ket{S}\otimes\ket{1}_{b}$ and
$\ket{1}=\ket{D}$ or $\ket{D}\otimes\ket{0}_{b}$, then one finds
\begin{eqnarray}
 R,R^{\dagger}:~~\alpha\ket{0}+\beta\ket{1}\to
(\alpha\ket{0}+\beta\ket{1})\cos\theta/2
+i (\alpha e^{-i\phi}\ket{1}+\beta
e^{i\phi}\ket{0})\sin\theta/2,
\end{eqnarray}
where $|\alpha|^{2}+|\beta|^{2}=1$.
\\
To generate W state it needs a sequence of 5 laser pulses
addressed to ion 2, 3 and 1
\begin{eqnarray*}
R^{+}_{2}(2
\arccos(1/\sqrt{3}),0)R_{3}(\pi,\pi)R_{3}^{+}(\pi/2,\pi)
R_{1}(\pi,0)R_{1} ^{+}(\pi,\pi),&&
\end{eqnarray*}
where first $R^{+}_{2}(2 \arccos(1/\sqrt{3}),0)$ is a
beamsplitter-like pulse on ion 2, which entangles its state with
the vibrational mode generating a non-symmetric superposition
$1/\sqrt{3}(\ket{SSS}\ket{0}_{b}+i\sqrt{2}\ket{SDS}\ket{1}_{b})$.
Next pulses result in the W state of ions
$1/\sqrt{3}(\ket{DDS}+\ket{DSD}+\ket{SDD}).$ In this experiment
for reconstruction of density matrix the state tomography has been
used, and fidelity of 83$\%$ was observed.
\\
The $^{40}Ca^{+}$ ion has an additional Zeeman level $D'$ so that
laser pulse on $S-D'$ transition can map the state
$\ket{S}\to\ket{D'}$ and back. The mapping allows readout
individual ion from the quantum ion trapped register while
preserving coherence. In the experiment with three-partite
entanglement the states of two ions were mapped into $S-D'$ space.
After reading the remainder ion, the laser pulses remap the states
into original space preserving coherence.
\\
The presented technics allows generating and manipulating
entanglement and are promising for quantum computing.

\subsection{Optical schemes for light}

\subsubsection{States of light and structure of the schemes}

 For quantum state engineering the optical implementation
of the $W$ states is attractive because of  set of simple
resources can be used. Light
 is usually presented by its modes, which  are specified by its wave
vectors $\mathbf{k}$  or "which path", polarization, say
horizontally $H$ and vertically $V$, and occupation number. So
$\ket{2H}_{\mathbf{k}}$ is a
 $H$-polarized mode with wave vector $\mathbf{k}$ and occupation number 2,
 but for shortness they often say about two $H$ photons, that pass
 along $\mathbf{k}$ direction or belong to the same space mode.
 In proposals and experiments two
 types of multimode states are discussed. First of them has one
 photon distributed into $n$ modes
 \begin{eqnarray}
\label{Wn1}
 W_{n}(1)=1/\sqrt{n}(\ket{1\dots 00}+\dots \ket{0\dots 01}),
 \end{eqnarray}
and second known as polarized $W$ state reads
\begin{eqnarray}
\label{WnV}
 W_{n}(V)=1/\sqrt{n}(\ket{V\dots HH}+\dots \ket{H\dots HV}).
\end{eqnarray}
The presented definitions are directly generalized to
$\eta_{n}(1)$ and $\eta_{n}(V)$.
\\
From point of  the view of QIT, that considers logical qubits,
both states (\ref{Wn1}) and (\ref{WnV}) are equivalent  up to
labelling $0\leftrightarrow H, 1\leftrightarrow V$, therefore they
have the same entropy, degree of entanglement and so on. Also
statistics of light in these states are similar. For $W_{n}(1)$
one finds the next correlation functions
\begin{eqnarray}
 \nonumber
\mean{a_{k}}=\mean{a_{k}a_{m}}=0,&&\\
\mean{a^{\dagger}_{k}a_{m}}=
\mean{a^{\dagger}_{k}a_{k}a^{\dagger}_{m}a_{m}}=1/n.&&
\end{eqnarray}
It follows, that each of the modes has  subpoissonian statistics
of photons with the Mandel parameter $\xi=-1/n$. In modes the
photons are anti-correlated because  the coincident rate of the
photon counting $\mean{a^{\dagger}_{k}a_{k}a^{\dagger}_{m}a_{m}}$
is less then product
$\mean{a^{\dagger}_{k}a_{k}}\mean{a^{\dagger}_{m}a_{m}}$. All
these properties are true for $W_{n}(V)$.
\\
There is a difference between these states. Indeed all current
proposals based on the linear lossless optical elements  permit
only conditional preparation of $W_{n}(V)$ in contrast to
$W_{n}(1)$.
\\
Considering the optical schemes proposed for generation of the $W$
states, one finds a similar structure of them. Their main
resources are linear optical elements $U$, sources of light $S$
and photon detectors $D$. A set of the linear elements is
presented mainly by beamsplitters ($BS$), polarized beamsplitters
($PBS$), half- and quarter-wave plates (HWP, QWP), and others.
These devices are passive, conserve the number of photons and can
be described by an unitary transformation $U$. One of the most
popular sources of light
 is the type-I and the type-II parametric down convertor ($PDC$)
 in the threshold regime known as spontaneous parametric source
 (SPDC), also the single-photon
source ($SPS$) is often discussed. For several experimental
proposals it requires commercial photon detectors, which can't
resolve the photon number of detection. So that these elements are
seem to be feasible by current technologies.
\\
Each scheme has two partitions at last. First splits the photons
of sources by linear optical elements into the output photons $O$
and the working photons $M$:
\begin{eqnarray}
  S\to U_{s} \to [O-M].
\end{eqnarray}
In the second partition after some unitary transformations all $M$
photons come to detectors and $O$ photons leave the schemes:
\begin{eqnarray}
    \leftarrow U_{O} \leftarrow  [O-M] \to U_{M} \to D.
\end{eqnarray}
The key idea is simple. A given set of linear optical elements
entangles photons of the source and distributes them so that their
superposition contains a desired state of $O$ photons. It is
extracted with a probability by a projection measurement on $M$
photons. The probability of the successful outcome depends on the
resources used and is one of the main characteristic of these
schemes.

\subsubsection{Experimental proposals}

The above abstract arguments can be found by examining several
experimental schemes proposed.
\\
First note, that a scheme for $W$ state using third order
nonlinearity for path entangled photons has been introduced by
Zeilinger et al. \cite{ZeilingerNASA Conf}.
\\
In the scheme for the generation of $W_{4}(V)$ introduced by
Mathis et al. \cite{Mathis} the presented setup consists of the
type-II $PDC$ and two $SPS's$ as input modes. With the help of the
post-selection strategy developed for  $GHZ$ \cite{Pan} the
$W_{4}(V)$ state can be achieved with probability 2/27 also
$W_{3}(V)$ can be done.
\\
An example of preparation of $W_{n}(1)$ with probability 1 and
$W_{3}(V),W_{4}(V)$ is given by Tomita et al. \cite{Bao-Sen Shi}.
The main resources are a set of $n$ single-photon sources and a
lossless $n\times n$ multiport fiber beamsplitter. If one photon
enters the input of the multiport beamsplitter the output state is
$\eta_{n}(1)$
 or particular $W_{n}(1)$ for symmetric configuration
 because of one photon is distributed with $n$ output
modes
\begin{eqnarray}
U_{n}:~~\ket{10\dots 0}\to W_{n}(1).
\end{eqnarray}
This method is generalized to $W_{n}(V)$ but only conditional
schemes can be achieved. One of the reasons of it is that already
two input photons are transformed by a beamsplitter into three
states $a\ket{11}+b\ket{20}+c\ket{02}$, however two of them $
\ket{20}, \ket{02}$ are often unwanted. Then it needs a
post-selection. Let three photons $H,H,$ and $V$ enter the input
of a tritter $U_{3}$ at the same time, then
\begin{eqnarray}
\label{Tritter}
 U_{3}:~~
\ket{1H}_{1}\ket{1H}_{2}\ket{1V}_{3}\to aW_{3}(V)+\dots.
\end{eqnarray}
If we select outcome in which there is only one photon in each
output then the polarized $W_{3}(V)$ states are  obtained with
probability 1/9. The found setup results in $W_{4}(V)$ with
probability 1/16, which is larger, than 2/27 of Ref. \cite{Mathis}
\\
Two schemes for $W_{3}(V)$ with  type-II PDC and a set of SPS's is
discussed by Yamomoto et al. \cite{Yamamoto}. In the first scheme
beamsplitters transform the four photon states of PDC into a
superposition of the form
\begin{eqnarray}
\ket{2H}_{\mathbf{k}}\ket{2V}_{\mathbf{k}}\to \alpha
\ket{1V}_{\mathbf{k}} W_{3}(V)+\ket{1H}_{\mathbf{k}} W_{3}(H)
\end{eqnarray}
Then after a projection measurement on the working photon in
$\mathbf{k}$ mode the $W_{3}(V)$ or $W_{3}(H)$ is prepared with
maximal probability 3/32. The second scheme includes three SPS's
and is similar to (\ref{Tritter}).
\\
Using schemes proposed by Kobayashi et al. in ref. \cite{Kobayashi
multiphoton entangled states} the $W_{4}(V)$, GHZ and ZSA states
are generated. The experimental setup includes type-I and type-II
PDC. In the first scheme the state of sources is transformed by a
tritter
\begin{eqnarray}
(\ket{1H}_{a}\ket{1V}_{b}+\ket{1V}_{a}\ket{1H}_{b})
\ket{2H}_{c}\to aW_{4}(V)+\dots.
\end{eqnarray}
and projects onto a single photon state. By this way $W_{3}(V)$
can be done with probability 0.0165. In this scheme the type-I PDC
can be replaced by  laser beam, from which the Fock state
$\ket{mH}$ originates. In the
 second scheme the initial state involved two photons from each PDC
$\ket{4H}_{a}\ket{0}_{b}+\ket{2H}_{a}\ket{1H1V}_{b}+\ket{0}_{a}
\ket{2H2V}_{b}$ are  transformed into $W_{4}(V)$ or to ZSA state.
Indeed, in these schemes all photons are working and are detected.
\\
Type-I PDC of a two-crystal geometry is proposed by Kobayashi for
generation of four-photon entanglement \cite{Kobayashi
two-crystal}.
\begin{figure}[ht]
\label{Bs}
  \centering
\epsfxsize=8cm \epsfbox{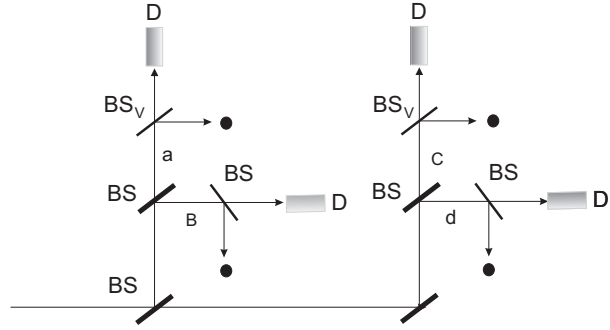}
  \caption{Scheme for generation of $W_{4}(V)$
   using four-photon states $\alpha\ket{4H}+\beta\ket{4V}+\gamma\ket{2H2V}$ of
   Type-I PDC of a two-crystal geometry}
\end{figure}
The scheme presented at fig. 1. The source produces the
four-photon state $a\ket{4H}+b\ket{4V}+c\ket{2H2V}$.  Experimental
setup includes beamsplitters, that transform the state of source
into entangled $a W_{4}(H)+b W_{4}(V)$ at a, b, c and d outputs.
To extract $W_{4}(V)$ there is a set of beamsplittes $BS_{v}$,
which have a small transparency for $V$ photons. By this way the
obtained state of the transmitted and reflected photons reads
$W_{4}(H)$ and $W_{4}(V)$ and is achieved by projective
measurement. Consider efficiency of these scheme. It can be
calculated assuming, that the probability of generation a photon
pair is about $\nu=4\times 10^{-4}$ per pulse. Probability of the
four-photon events the probability has order of $\nu^{2}=8\cdot
10^{-8}$. For a pump laser with a 100-MHz repetition rate the
generation rate of $W_{4}(V)$ is about $10^{-1}$ per second and
with the 50\% detection efficiency one finds 1 desired state per
minute.

\subsection{Experiments}

The recent experiments for generating multiphoton entanglement are
based on PDC and linear optics elements manipulating polarized
light.
\\
Weinfurter et al have proposed a type-II PDC
\cite{WeinfurterZukowski} that has been used as a source in many
experiments. Its state is a superposition of the four-photon $GHZ$
and the tensor product of two maximally entangled EPR pairs,
emitted into the two spatial modes
\begin{eqnarray}
\Psi^{(4)}=\sqrt{2/3}\ket{GHZ}-\sqrt{1/3}\ket{EPR}
 \ket{EPR}.
\end{eqnarray}
Here $\ket{EPR}=(\ket{H}_{a}\ket{V}_{b}+
\ket{V}_{a}\ket{H}_{b})/\sqrt{2}$, and $GHZ$ has two
indistinguishable photons of the same polarization into one space
mode : $\ket{GHZ}=(\ket{2H}_{a}\ket{2H}_{b}+
\ket{2V}_{a}\ket{2V}_{b})/\sqrt{2}$. This is not a product of two
entangled pairs. If each of  two modes splits at a beamsplitters,
correlations between four photons can be observed \cite{Eibl
four-photon}. However it needs to select events such that one
photon is detected in each of the four modes. Two types of
coincidence due from $GHZ$ and $EPR$ state between 300 and 100 per
hour for integration time of 5 and 17.5 h. have been observed. By
this way the Bell inequality for four qubits has been tested.
\\
Using the source generated $\Psi^{(4)}$, Weinfurter et al
\cite{EiblThreeQubit} have generated the $W_{3}(V)$ state and
examined its entanglement. The experimental setup is given at fig.
2.
\begin{figure}[ht]
\label{Bs}
  \centering
\epsfxsize=8cm \epsfbox{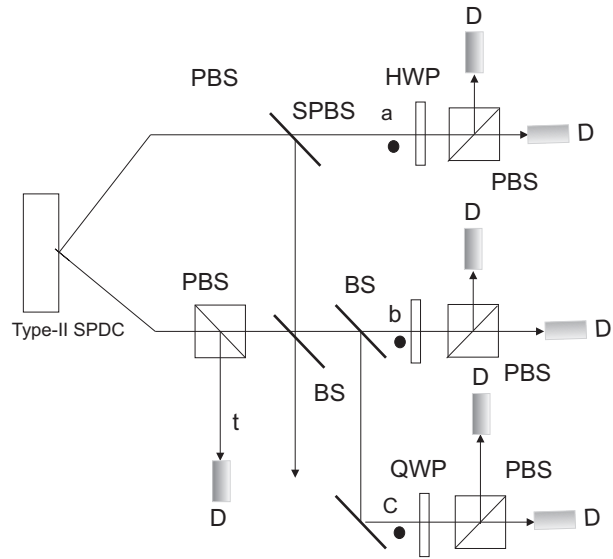}
  \caption{Experimental setup to generate $W_{3}(V)$ by the SPDS source
  state of $\Psi^{(4)}=\sqrt{2/3}\ket{GHZ}-\sqrt{1/3}\ket{EPR}
 \ket{EPR}$ }
\end{figure}
Four photons in the $\Psi^{(4)}$ state are distributed into four
modes $t,a,b,$ and $c$. By detection a  photon in each of the arms
the $W_{3}(V)$ entanglement in $a,b,c$ modes is prepared. In
experiment several characteristics have been tested to verify the
observed state. First is the coherence arisen from of the wave
function. The reason is that the desired state is produced by
independent measurements of the modes and coincidence
photocurrents are examined. However by this way coherence of the
wave function can't be detected and both the pure state and mixed
one can't be distinguished. Second is the robustness of the $W$.
In experiment it has been tested by performing an  one-mode
measurement, that projects a  mode in to $H$ or $V$ state. It
needs two measurements of the such type to destroy entanglement.
So that after projecting in $\ket{H}_{a}$ an EPR pair of the form
$1/\sqrt{2}(\ket{HV}+\ket{VH})_{bc}$ has been observed. Also the
generalized Mermin inequalities \cite{Mermin} has been examined.
However Cereceda has pointed out, that these inequalities can't
verify the tripartite entanglement \cite{Cereceda}
\\
A problem of detection of genuine multiparticle entanglement of
$W_{3}(V)$ and $\Psi^{(4)}$  has been studied experimentally by H.
Weinfurter et al. \cite{Weifurter Witness}.
 A set of witness
operators has been measured for $W$ and $\Psi^{(4)}$ using a set
of polarization analyzers which consist of QWP, HWP and PBS. One
of the witness for $W$ is given by (\ref{Ww1}), where the spin
observable $\sigma_{z}$ corresponds to measurement of $H$, $V$
linear polarization, $\sigma_{x},\sigma_{y}$ corresponds to
analysis of $\pm 45^{0}$ linear polarization (left-right circular
polarization). By this way the genuine  $W$ entanglement has been
demonstrated. Indeed for four -photon state, 15 different analyzer
settings are required.
\\
Several quantum informational tasks can be done using the
four-photon state $\Psi^{(4)}$. Weifurter et al. \cite{Weifurter
DF} have demonstrated preparation of decoherence-free states which
enable to encode a qubit in decoherence-free space. The scheme is
presented at fig. 3.
\begin{figure}[ht]
\label{Bs}
  \centering
\epsfxsize=10cm \epsfbox{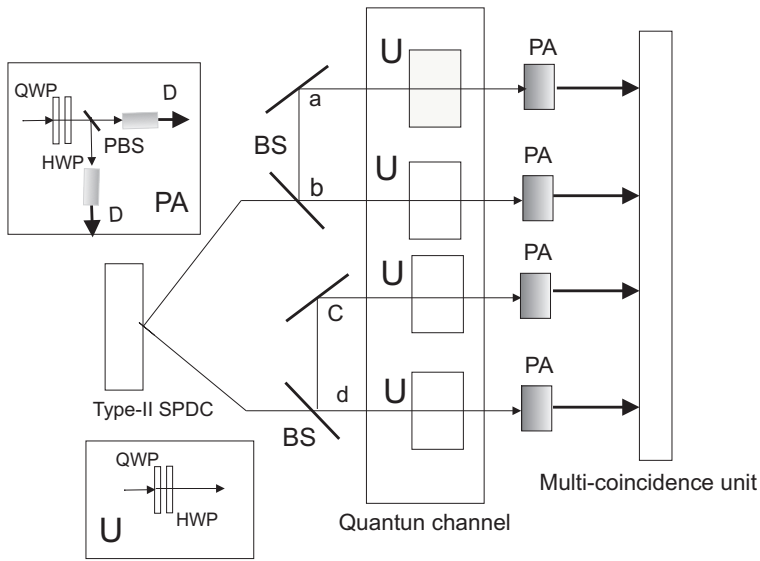}
  \caption{Experimental setup for decoherence-free communication.
  The photons of the source are distributed into four mode a,b,c,d.
  The noisy channel is presented by quarter- (QWP) and half wave
  plates (HWP). Polarization analyzers (PA) are employed for the registration
  of the decoherence-free states.}
\end{figure}
 In experiment two DF states of the form
(\ref{DF1}) and (\ref{DF2}) were generated from $\Psi^{(4)}$ and
sent into noisy channel. The channel is presented by inserting QWP
and HWP in each modes. The invariance of the encoded information
has been observed by comparing the density matrix before and after
the interaction with environment.
\section{Applications}
Some  information tasks need entanglement so that there is a
question whether MES of the W-class can be used. It has been found
that some of these states are suitable for the problem of secrete
sharing  and key destribution \cite{Jaewoo JooSS}, teleportation
\cite{Jaewoo JooTel} and dense coding also the distillation
protocols \cite{Zhuo-Liang Cao Ming Yang1}, \cite{Zhuo-Liang Cao
Ming Yang2} have been proposed.
\subsection{Quantum key distribution and secrete sharing} Key
distribution and secrete sharing are problems of  classical
cryptography and can be implemented using  quantum resources,
particularly $W$ state and projective measurements. Let three
parties Alice, Bob and Claire share the   $W$ state given by
(\ref{W}) and perform randomly measurement of $\sigma_{x}$ and
$\sigma_{z}$ on his own particle. The key idea of using $W$ state
is that after projecting onto $\ket{0}_{A}$ Bob and Claire have in
their hands entanglement $(\ket{01}+\ket{10})/\sqrt{2}$ and the
subsequent outcomes of their measurements will be correlated. In
contrast after projecting onto $\ket{1}_{A}$ they have a product
$\ket{00}$ and  the independent outcomes aries. This is a basis of
quantum key distribution (QKD) because  two correlated outcomes
represent a key bit in the Bob and Claire hands. Quantum secrete
sharing (QSS) is a QKD among $n$ persons in such a way that one's
key message can be retrieved by the $k\leq n$ persons if they
cooperate all together.
\\
In more detail. To describe the measurement of $\sigma_{x}$ and
$\sigma_{z}$ introduce their eigenvectors
\begin{eqnarray}
\sigma_{x}\ket{x_{\pm}}=x_{\pm}\ket{x_{\pm}},~~~\sigma_{z}\ket{z_{\pm}}=z_{\pm}\ket{z_{\pm}},&&
\end{eqnarray}
where $\ket{x_{\pm}}=(\ket{0}\pm\ket{1})/\sqrt{2},
\ket{z_{\pm}}=\ket{0},\ket{1}$, and $x_{\pm},z_{\pm}=\pm 1$ are
eigenvalues or outcomes of the measurement. In the case of
projecting, for example, into $\ket{z_{+}}$ outcome of the
measurement is $z_{+}$. Using these notations $W$ state can be
written in the form
\begin{eqnarray}
\nonumber
W=(1/\sqrt{3})[\ket{z_{+}z_{-}z_{-}}+\ket{z_{-}z_{+}z_{-}}
\ket{z_{-}z_{-}z_{+}}]&&\\
=(1/\sqrt{3})\ket{z_{+}}[\ket{x_{+}x_{+}}-\ket{x_{-}x_{-}}]
+(1/2\sqrt{3})\ket{z_{-}}[\ket{x_{+}x_{+}}+\ket{x_{+}x_{-}}+
\ket{x_{-}x_{+}}+\ket{x_{-}x_{-}}].&&
\end{eqnarray}
 When Alice has outcome $z_{+}$ then Bob and Claire obtain
the same outcome $x_{+}$ or $x_{-}$ which is a key bit.
 The success probability in distributing a key bit is 2/3 .
The case when Alice has outcome $z_{-}$  is useless and discarded.
Note, the difference in correlation between two states
$\Psi^{+}=(\ket{01}+\ket{10})/\sqrt{2}$  and $\ket{00}$ in the Bob
and Claire hands. It is presented here by the number of outcomes,
which are $x_{+}x_{+};~ x_{-}x_{-}$ for $\Psi^{+}$ and
$x_{+}x_{+};~ x_{+}x_{-};~x_{-}x_{+};~ x_{-}x_{-}$ for $\ket{00}$.
So any correlation may reduce the total number of outcomes. The
protocol requires 12 qubits per a key bit at average. On the other
hand the protocol E91 (based on Bell's theorem and EPR pairs as  a
quantum channel) has the overall success probability 2/9 and it
requires 9 qubits per bit.
\\
In QSS  Bob  and Claire  are expected to retrieve message from
Alice in their cooperation. If Alice has outcome  $z_{+}$ then Bob
and Claire have opposite outcomes out of $z_{+}$ and  $z_{-}$.
Otherwise both have the same outcome $z_{+}$.  When Bob and Claire
cooperate they can  collect their outcomes  to correctly deduct
the key bit of A. The overall success probability is 1/8 which is
determined by the probability of the choosing measurements.  Due
to this argument 24 qubits are necessary  to share a key bit.
It has been shown that these protocols are secure against simple
individual attacks by an eavesdropper. These attacks are such that
Eve performs an unitary operation on a composite system of her
auxiliary qubit and one of the three qubits which are involved in
a secure communication and she tries to extract some information
by measuring her auxiliary qubit \cite{Gisin}.

\subsection{Distillation of W}
When an entangled state is transmitted its quantum correlations
can be destroyed because of noise. To achieve a faithful
transmission Bennet at all  have proposed purification of the
state using LOCC \cite{Bennett Purification}. It can be done using
a set of the Pauli and CNOT operations. In \cite{Zhuo-Liang Cao
Ming Yang2} a protocol for distillation of the form
\begin{eqnarray}
 a\ket{100}+b\ket{010}+c\ket{001}\to (1/\sqrt{3})
 (\ket{100}+\ket{010}+\ket{001})
\end{eqnarray}
has been proposed. It includes two points: 1/ unitary
transformations $U_{k}, k=1,2$ on $W$ and ancilla qubit
\begin{eqnarray}
U_{k}=
\begin{pmatrix}
  1 & 0 & 0 & 0 \\
  0 & v_{k} & 0 & \sqrt{1-v^{2}_{k}} \\
  0 & 0 &-1 & 0 \\
  0 & \sqrt{1-v^{2}_{k}} & 0 & -v_{k} \\
\end{pmatrix},
\end{eqnarray}
where $v_{1}=c/a$ and $v_{2}=c/b$, 2/ a measurement on ancilla.
The successful probability is $3c^{2}$. This protocol can be
simulated in cavity QED.

\subsection{Teleportation and Dense coding using W-channel}

Quantum teleportation, that  allows transmitting an unknown
states, is attractive for communications also for computing as
primitive for quantum computations  \cite{Gottesman}. The protocol
has been demonstrated experimentally for teleportation of
polarized photon \cite {D.Bouwmeester}, coherent state of light
\cite{Furusawa} and atom \cite{AtomTeleport}. In the standard
protocol an unknown qubit state
\begin{eqnarray}
\varphi=\alpha\ket{0}+\beta\ket{1}
\end{eqnarray}
can be transmitted using an $EPR$ pair as a quantum channel,  2
bits of classical information gained in the Bell-state measurement
and Pauli matrixes, which are retrieval operators.
\\
Instead of EPR pair Karlsson et al. \cite{Karlsson} have
considered the tree-particle $GHZ$ entanglement for sending
unknown qubit to two receivers. It can be done probabilistically
because of the non-cloning theorem, which forbids copying of an
unknown states by linear unitary transformations \cite{Zurek}.
However one of the receivers can retrieve unknown state if he'll
cooperates with other receiver.
\\
Can the $W$ states be suitable  for teleportation as a quantum
channel? In a large number of the presented protocols the task of
transmitting unknown qubit is accomplished probabilistically only
\cite{ShiTomita}, \cite{Jaewoo JooTel}. Nevertheless there is an
unconditional protocol for teleportation of entangled state
\begin{eqnarray}
\label{Ent}
 \phi=\alpha\ket{01}+\beta\ket{10}
\end{eqnarray}
which has been proposed in our Ref. \cite {GorbachevTrubilko}. It
based on the observable, that two entangled qubits can be
transmitted perfectly in the $GHZ$-channel by 3 bits of classical
information if a Bell-like state measurement \cite{Luca} is
performed
\begin{eqnarray}\label{GHZTel}
\phi_{12} \otimes |GHZ\rangle_{ABC}
=(1/\sqrt{8})\sum_{x}|\Phi_{x}\rangle_{12A}[B_{x}\otimes
C_{x}]\phi_{BC},
\end{eqnarray}
where each of the eight vectors $\Phi_{x}$ is a product of the
Bell state and eigenvector of $\sigma_{x}$, the retrieval
operators $B$, $C$ are defined by Pauli matrixes. This equation
tells, that if the GHZ channel allows teleportation of a state
$\phi$, then this state can be teleported using any channel,
obtained from the GHZ one by unitary transformation, that involves
all particles of the channel except one.
\\
The required two-particle transformation reads
\begin{eqnarray}\label{T203}
V=|\Psi^{+}\rangle\langle 00|+ |11\rangle\langle 01| 
+|\Psi^{-}\rangle\langle 10|+|00\rangle\langle 11|.&&
\end{eqnarray}
It is a non-local unitary operation, that convert $GHZ$ into a
state from the $W$ class
\begin{eqnarray}\label{T2030} (1\otimes
V)|GHZ\rangle_{ABC}=
(1/\sqrt{2})\ket{100}+(1/2)\ket{010}+(1/2)\ket{001}.
\end{eqnarray}
By applying this transformation to both sides of (\ref{GHZTel}) we
have teleportation of entangled state by the channel of the $W$
class. There is new feature of the recovering operators which
become non-local. The obtained $W$ channel can accomplish
tree-qubit dense coding, when three bit of classical information
can be send by manipulating two qubits \cite{3QubitC},
\cite{GTRZ}.
\section{Conclusions}
Considering MES of the $W$-class we find interesting properties,
proposals of their implementations and experimental realizations.
They can be used for teleportation and dense coding also in
quantum quantum cryptography for key distribution, secrete sharing
and others. One of the important features of these states, that
follows from their entanglement, is robustness, which
distinguishes them from another states. So it has been shown that
they have immunity to decoherence being decoherence-free states.
However exploiting of this properties is not easy problem and we
think that one of the main open questions is how to use fully the
potential of MES from the $W$ class.
\\
\section*{Acknowledgments}
We are grateful to A.M. Basharov for hopeful discussion. This work
was supported in part by the Delzell Foundation Inc.

\end{document}